\documentclass[showpacs,showkeys]{revtex4}%
\usepackage[colorlinks=true]{hyperref}

\AtBeginDocument{
	\hypersetup{
		urlcolor=cyan,
		citecolor=magenta,
		linkcolor=orange,
		anchorcolor=green,
	}%
}
\usepackage{amssymb}
\usepackage{amsmath}
\usepackage{amsfonts}
\usepackage{graphicx}
\usepackage{refcheck}
\usepackage{ulem} 

\usepackage{cancel}
\usepackage[usenames]{xcolor}
\usepackage{epstopdf}
\usepackage{extarrows}
\providecommand{\sqbr}[1]{\left[ #1 \right]} %
\providecommand{\ttiny}[1]{\text{\tiny{#1}}}%

\def\Ra{\Rightarrow}

\begin{document}

\title[ ]{Impossible Mission: Entropy Maximization with Escort Averages}
\author{$^{1}$Thomas Oikonomou}
\email{thomas.oikonomou@nu.edu.kz}
\author{$^{2}$G. Baris Bagci}
\affiliation{$^{1}$Department of Physics, School of Science and
Technology, Nazarbayev University, Astana
010000, Kazakhstan}
\affiliation{$^{2}$Department of Materials Science and Nanotechnology Engineering, TOBB University of Economics and Technology, 06560 Ankara, Turkey}

\begin{abstract}

It has recently been a common practice to maximize the deformed entropies through the escort averaging scheme. However, whatever averaging procedure is employed, one should recover the ordinary Shannon maximization results in the appropriate limit of the deformation parameter e.g. $q\to1$ for the Tsallis and R\'enyi entropies. Otherwise, the very meaning of a consistent generalization becomes at stake. Using only this equivalence, we show that any deformed entropy expression, maximized with the escort averaged constraints, yields that the Shannon entropy $S$ is equal to the logarithm of the ordinary canonical partition function i.e. $S=\ln(Z_\ttiny{S})$ instead of the correct thermodynamic relation $S=\beta U+\ln(Z_\ttiny{S})$. Therefore, we conclude that the use of the escort averaging procedure should be avoided for any deformed entropies, since it cannot even yield the well-known thermodynamic relations of the ordinary canonical formalism.       
\end{abstract}

\eid{ }
\date{\today }
\startpage{1}
\endpage{1}
\maketitle

\section{Introduction}
%

The Maximum Entropy (MaxEnt) principle \cite{Jaynes1} relies on the maximization of the entropy measure $S$ with some appropriate constraints. As a result of this procedure, MaxEnt picks the probability distribution  yielding the highest entropy $S$ among all the other possible distributions conforming to the pre-chosen constraints.

Although MaxEnt procedure has been originally used for the Shannon entropy, it has also been a common tool for the recently developed generalized deformed entropies such as the Tsallis \cite{Tsallis1988} and R\'enyi \cite{Renyi} entropies which found numerous applications in various fields of research \cite{Bagci1,Rajagopal,Rotundo,Bagci2,Van1,Van2,Wong,Chang,reis,mendes}. However, there is no consensus on how the constraints, in particular the internal energy constraint, should be incorporated into the MaxEnt procedure in the case of the generalized entropies \cite{Bashkirov,TsallisMP,Bagci3,Oik2007,Lenzi}. The choice of the proper constraints and the concomitant averaging procedure is not only an academic issue without practical interest, since it is closely related to the third law of thermodynamics \cite{Bagci4,Bagci5}, the orthode approach of Boltzmann \cite{Campisi} and the stability considerations necessary for a feasible thermodynamics \cite{stab1,stab2,stab3}. We also note that different alternative approaches to MaxEnt have been developed for the generalized entropies \cite{PlastPlastCurado,PlastinoCurado,miller}.

One choice of constraints is formed by considering the internal energy written in terms of the escort probabilities used in many applications so far \cite{escortuse}. This choice reproduces the linear averaged internal energy constraint in the $q\to1$ limit where $q$ is the deformation parameter. Since the deformed entropies such as Tsallis and R\'enyi entropies become the Shannon entropy in this limit, one should exactly recover the results obtained from (i.e. after) the Shannon entropy maximization procedure with linearly averaged internal energy constraint in this particular limit. In fact, this equivalence between the deformed entropies and the Shannon entropy in the $q\to1$ limit is generally taken for granted and used in order to check many equations obtained in the framework of deformed entropies $q\to1$.        

In the next section, using this equivalence, we show that the use of escort averages incorrectly implies that the Shannon entropy is equal to the logarithm of the canonical partition function. Note that the calculation of this section is valid for any generalized entropy converging the Shannon measure in the $q\to1$ limit. The concluding remarks are presented in section III.

\section{MaxEnt and Deformed Entropies: General Discussion}\label{MaxEntF_1}
%
%
In the canonical ensemble the MaxEnt approach for an arbitrary deformed entropy $S_q(\{p\})$ \cite{DefLog}, where $q$ is the deformation parameter, with the escort energy averaging procedure
\begin{eqnarray}\label{EscMV}
U_q:=\frac{\sum_{i=1}^{n} p_i^{q}\varepsilon_i}{\sum_{k=1}^{n}p_k^{q}}\,,
\end{eqnarray}
relies on the following functional 
\begin{eqnarray}\label{ExtFun}
\Lambda=S_q
-\alpha \sqbr{\sum_{i=1}^{n} p_i-1}
-\beta \sqbr{\frac{\sum_{i=1}^{n} p_i^{q}\varepsilon_i}{\sum_{k=1}^{n}p_k^{q}}-U_q}\,,
\end{eqnarray}
where $p_i$ is the micro-probability, $\varepsilon_i$  and $U_q$ denote the micro-energy and the internal energy, respectively. For $q\to1$ limit, both the escort average and the entropy $S_q$ recover the  linear mean average and the Shannon entropy $S$, respectively. As usual,  $\alpha$ and $\beta$ are the so-called Lagrange multipliers associated with the normalization  and the internal energy constraints. Particularly, labelling $\alpha$ after the maximization of Eq. (\ref{ExtFun}) as $\alpha_q$, we see that
\begin{eqnarray}
\alpha_q=\sum_{i=1}^{n}p_i\frac{\partial S_q}{\partial p_i}\,.
\end{eqnarray}
On the other hand, it is easy to check that for the Shannon case after the maximization with the linear mean averaging procedure i.e.  $\alpha=\alpha_\ttiny{S}$, one obtains the following expression
\begin{eqnarray}
\alpha_\ttiny{S}=\ln(Z_\ttiny{S})-1\,,
\end{eqnarray}
where $Z_\ttiny{S}$ is the canonical partition function,
\begin{eqnarray}
Z_\ttiny{S}=\sum_{i=1}^{n} e^{-\beta \varepsilon_i}\,.
\end{eqnarray}
Accordingly, as a matter of consistency, when $q\to1$, the normalization multiplier $\alpha_1$ must recover $\alpha_\ttiny{S}$ after the maximization procedure. However, when we equate the two i.e. $\alpha_1=\alpha_\ttiny{S}$, we obtain
\begin{eqnarray}
\sum_{i=1}^{n}p_i \frac{\partial S}{\partial p_i}=\ln(Z_\ttiny{S})-1 \quad\Ra\quad
\sum_{i=1}^{n}p_i\sqbr{\ln(1/p_i)-1}=\ln(Z_\ttiny{S})-1
\end{eqnarray}
which immediately yields

\begin{eqnarray}
S=\ln(Z_\ttiny{S}).
\end{eqnarray}

In other words, one always relies on the equivalence between the escort averaging scheme in the $q\to1$ limit and the ordinary Shannon entropy maximization. However, this assumed equivalence implies that the the Shannon entropy is just the logarithm of the ordinary canonical partition function $S=\ln(Z_\ttiny{S})$ instead of the well-known relation $S=\beta U+\ln(Z_\ttiny{S})$.

\section{Conclusions}

In maximizing the deformed entropy expressions, the so-called escort averaged internal energy constraint i.e. $U_q= \sum_{i=1}^{n} \frac{p_i^{q}}{\sum_{k=1}^{n}p_k^{q}} \varepsilon_i $ is usually used instead of the more familiar expression $U= \sum_{i=1}^{n} p_i \varepsilon_i $ employed in the Shannon maximization procedure. Since the escort averaged internal energy constraint (and also the deformed entropy measures) becomes the linear averaged one in the $q\to1$ limit, one should recover the results of the ordinary Shannon maximization in the aforementioned limit. However, when checked, this simple observation yields the incorrect relation for the Shannon entropy i.e. $S=\ln(Z_\ttiny{S})$ instead of the well-known correct one, that is, $S=\beta U+\ln(Z_\ttiny{S})$. In this sense, the escort averages do not even yield the correct thermodynamic relations for the Shannon entropy in the appropriate limits. Therefore, the use of such averages for the deformed entropies should be avoided for a consistent thermodynamics.


\end{document}